\documentclass[10pt,letterpaper,twocolumn]{article} 

\usepackage{ol2}
\usepackage[draft]{hyperref}
\usepackage{amsmath,braket}

\begin{document}

\twocolumn[ 

\title{Measurement of two-mode squeezing with photon number resolving multi-pixel detectors}


\author{Dmitry A. Kalashnikov,$^1$ Si-Hui Tan,$^1$ Timur Sh. Iskhakov,$^2$ Maria V. Chekhova,$^{2,3}$ \\and Leonid A. Krivitsky$^{1,*}$}
\address{
$^1$Data Storage Institute, Agency for Science Technology and Research (A-STAR), 5 Engineering Drive 1, 117608 Singapore
\\
$^2$Max Planck Institute for the Science of Light, G\"{u}nter Scharowski str., Bau 19, 91058 Erlangen, Germany\\
$^3$Department of Physics, Moscow State University, Leninskie Gory 1-2, 119992 Moscow, Russia\\
$^*$Corresponding author: Leonid\_Krivitskiy@dsi.a-star.edu.sg
}

\begin{abstract}The measurement of the two-mode squeezed vacuum generated in an optical parametric amplifier (OPA) was performed with photon number resolving Multi-Pixel Photon Counters (MPPCs). Implementation of the MPPCs allows for the observation of noise reduction in a broad dynamic range of the OPA gain, which is inaccessible with standard single photon avalanche photodetectors.\end{abstract}

 ] 

\noindent Multiphoton entangled states attract considerable attention in modern quantum optics, as they represent promising resources for quantum communications, quantum computing, quantum lithography, and quantum imaging \cite{klm, kok}. One of the most accessible ways to generate such states is by using parametric down conversion (PDC), a non-linear process producing pairs of strongly correlated photons. The state produced by frequency non-degenerate PDC is a two-mode squeezed vacuum (SV) state written as follows \cite{Klyshko}:
\vspace{-2 mm}
\begin{eqnarray}
\ket{\psi}=\sum^\infty_{n=0}C_n \ket{n}_s\ket{n}_i \ ,
\end{eqnarray}
where $\ket{n}_j$ denotes the Fock-state of $n$ photons in the $j$-th mode, $j=s,i$ denote signal and idler modes respectively, and $C_n$ is the probability amplitude. Strong correlation between the photon numbers in the signal and idler modes, referred to as two-mode squeezing, results in the suppression of the variance of their difference below the shot-noise limit \cite{Masha, Fabre} and can be quantified by the noise reduction factor (NRF) given by
\vspace{-2 mm}
\begin{eqnarray}\label{NRF}
NRF=\frac{{\rm Var}(\widehat{N}_s-\widehat{N}_i)}{\braket{\widehat{N}_s+\widehat{N}_i}} \ ,
\end{eqnarray}
where $\widehat{N}_s$ and $\widehat{N}_i$ are the photocount numbers of the 
detectors in the signal and idler modes respectively, and $\braket
{\widehat{O}}={\rm tr}(\hat{\rho}\widehat{O})$ defines the mean value of 
an observable $\widehat{O}$ for a given state $\hat{\rho}$. The two-mode 
squeezing can be studied with various types of photodetectors, depending 
on the photon fluxes of the PDC. At a moderate gain, when the SV state 
contains several photons per pulse, it is essential that the detector is 
able to resolve several simultaneously impinging photons. Such detectors 
are referred to as photon number resolving detectors (PNRDs). 

Development of the PNRDs attracts a considerable interest in modern 
optical engineering. Several PNRD technologies are available up to date, 
including cryogenic detectors \cite{Hogue, Waks, Nam}, branched single 
photon avalanche detectors (SPADs) \cite{Walmsley,Franson}, hybrid 
photodetectors \cite{Bondani1}, intensified and electron-multiplied CCDs 
\cite{Lantz}. 

Here we consider a multi-pixel photon counter (MPPC) which is a 
commercially accessible PNRD \cite{Hamamatsu}. In a MPPC, several hundreds 
of SPADs (pixels) are embedded into a chip which is illuminated by a 
diffused light beam, such that the chance of more than one photon hitting 
the same pixel is negligible. Upon photon detection, a pixel produces an 
electrical signal, which is summed with the signals from other pixels at 
an output circuit, thus enabling the photon number resolution. Despite its 
attractive features, such as ease of operation, low cost, and high 
quantum efficiency (q.e.), the crucial drawback of the MPPC is the optical 
crosstalk between pixels. During an avalanche in a pixel, some spurious 
photons are emitted, and eventually registered by the neighboring pixels 
\cite{Rech}. As the crosstalk happens almost simultaneously with the 
detection of real photons, its contribution cannot be distinguished from 
the actual counts, and therefore requires an accurate theoretical modeling 
\cite{Rech,Afek, Gisin}.

The performance of the MPPC has been studied in 
earlier experiments with coherent and pseudo-thermal light \cite{Rech, 
Afek, Gisin, SPIE, Ramilli}. In \cite{Dima1} the MPPC was used in 
the measurements of the intensity correlation function of the SV state, 
however restricted to the regime of much less than one photon per pulse. 
In this work we expand the application of the MPPC to the study of 
a relatively bright SV state with up to almost 5 photons per pulse by the 
direct observation of the two-mode squeezing.

For modeling a realistic MPPC, we follow \cite{Afek}, where loss and 
crosstalk are considered as Bayesian processes. Let us denote $\braket
{\hat{n}}$ as the mean photon number impinging on the detector, $P$ as the 
crosstalk probability, and $\eta$ as the q.e. of a single SPAD pixel, which also accounts for the losses in the optical system. Here, beams of relatively high intensities have been considered hence the saturation of the MPPCs has to be taken into account by assuming that each MPPC can only resolve up to $N_{\rm max}$ photocounts. We assume that the two MPPCs used in joint detection have the same values of $\eta$, $P$, and $N_{\rm 
max}$. Then the positive-operator valued measure for detecting $N_j
$ photocounts in the $j$-th mode is given by
\vspace{-2 mm}
\begin{eqnarray}
\Pi_{N_j}=\left\{\begin{array}{cc}\displaystyle\sum_{n=\lceil \frac{N_j}{2} \rceil}^{N_j}\displaystyle\sum^
\infty_{k=n}B_{n,k,N_j}\ket{k}\bra{k} \ &, \ N_j<N_{\rm max} \\[1 ex] I-\displaystyle\sum_{k=0}^{N_
{\rm max}-1}\Pi_k \ &, \ N_j=N_{\rm max}, \end{array}\right.
\end{eqnarray}
where
\vspace{-2 mm}
\begin{eqnarray}
B_{n,k, N_j}(\eta, P)&=&\left(\begin{array}{c} n \\ N_j - n\end{array}
\right)\left(\begin{array}{c} k \\ n\end{array}\right)
\\
\nonumber &&P^{N_j-n}(1-P)^{2n-N_j}\eta^n(1-\eta)^{k-n}\ . \end{eqnarray}
The variance of the difference of photocount numbers is given by 
\cite{Agliati}:
\vspace{-2 mm}
\begin{eqnarray}
\label{second}{\rm Var}(\widehat{N_s}-\widehat{N_i})&=&\braket{\widehat{N_s^2}}-\braket{\widehat{N_s}}^2+\braket{\widehat{N_i^2}}-\braket{\widehat{N_i}}^2\\ \nonumber &&-2\braket{\widehat{N_s}\widehat{N_i}}+2\braket{\widehat{N_s}}\braket{\widehat{N_i}}\ ,
\end{eqnarray}
where the p-moment operator of the photocount number operator is given by $\widehat{N_j^p}=\sum^{N_{\rm max}}_{N_j=0}N_j^p \Pi_{N_j}$. Using (\ref{NRF}) -
(\ref{second}), the NRF can be calculated for any given two-mode state. 
As $\braket{\hat{n}} \rightarrow0$ the model yields $NRF=\frac{1+3P}
{1+P}$ for the coherent state which is greater than unity--the value 
expected for a lossy PNRD without the crosstalk. Thus, the effect of the 
crosstalk is to increase the NRF. Also, as $\braket{\hat{n}}
\rightarrow0$ for the SV state the model yields $NRF=\frac{1+3P}{1+P}-
(1+P)\eta$. Thus the difference in NRF values for the two states is 
equal to the effective q.e. of the detectors $\eta_E\equiv(1+P)\eta$, differing from the absolute q.e. due to the presence of crosstalk \cite{Agafonov,Dima2}. Numerical calculations show that the  effect of saturation is 
to decrease the NRF with increasing $\braket{\hat{n}}$ (see Fig. 2).

In the experiment the 4-th harmonic of Nd:YAG pulsed laser at 266 nm (Crystalaser, repetition rate 20 kHz, 30 ns pulse width) was used as a pump for a traveling wave OPO (Fig.1). It was sent through two 5 mm length BBO crystals with oppositely directed optical axes in order to compensate for the spatial walk-off of the pump. The BBOs were cut for collinear frequency non-degenerate type-I PDC with signal and idler wavelengths $\lambda_s=500$ nm and $\lambda_i=568$ nm, respectively. A half-wave plate and a polarization beamsplitter (not shown) were used to control the pump power. After the BBO crystals, a UV-mirror filtered out the pump beam, while the PDC beam was transmitted. The signal and idler beams were split by a dichroic mirror; the former was reflected and the latter was transmitted. In each arm there were lenses with their front foci at the BBOs. Two iris apertures were set according to $d_s/d_i=\lambda_s/\lambda_i$ in the back focal planes of the lenses, selecting the conjugated spatial modes \cite{Masha}. The spectral bandwidth of PDC was calculated to be 3 nm, and was restricted by the spatial selection. The PDC radiation was then collected by lenses and formed a beam spot of 1.2 mm (at level $1/e^2$) at the MPPC chip. The ambient light was suppressed by means of interference filters centered at 500 nm and 568 nm (15 nm FWHM, peak transmission 95$\%$, not shown). A 0.1 dB neutral density filter was introduced in the signal arm to balance the q.e. of the MPPCs to the value of the q.e. at 568 nm. Two MPPC modules (Hamamatsu, model 10751-02, 20$\times$20 pixels each of active area $50\times 50 \ \mu m^2$, total chip size $1.5\times1.5\ mm^2$), were sealed into a custom housing with an AR-coated optical window and were thermoelectrically cooled to -4.5$^{\circ}{\rm C}$ to reduce the dark noise. The signals from the MPPCs were digitized by an AD card (National Instruments, PCI-5154) and discriminated by their amplitudes. The AD card was synchronized with the pump laser with the detection window set to 70 ns. The data analysis was done with LabView, and the theoretical fits with Mathematica. The MPPC dark noise and ambient light of the setup were measured with an extinguished PDC, for which the polarization of the pump was 90 degrees rotated by the half-wave plate, and then subtracted from the data.
\vspace{-2 mm}
\begin{figure}[htb]
\centerline{\includegraphics[width=7.5cm]{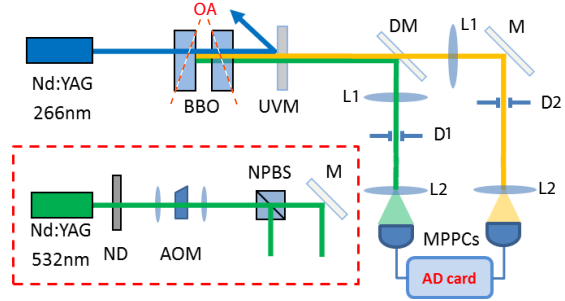}}
\caption{\small (Color online) The experimental setup: Nd:YAG laser at 266 nm pumps two BBO crystals. OA, optical axis; UVM, UV-mirror; DM, dichroic mirror; M, mirror; L1 (f=500 mm), L2 (f=200 mm), lenses; D1 ($d_s=10$ mm), D2 ($d_i$=11.4 mm), iris apertures. The inset shows the setup for measurements of the coherent state: cw Nd:YAG laser at 532 nm is chopped by an acousto-optical modulator (AOM); ND, neutral density filter; NPBS, non-polarizing beamsplitter.}
\end{figure}

Similar measurements were performed for the coherent light obtained by attenuating the 2-nd harmonic of a continuous wave (cw) Nd:YAG laser at 532 nm (Photop, power 50 mW) and chopping it by an acousto-optical modulator at a frequency of 20 kHz with a pulse duration of 30 ns.

Fig. 2 shows the experimental data and theoretical fits of NRF versus $\braket{\hat{n}}$ for 
the coherent and SV states. From the difference of the measured NRFs for the two states close to $\braket{\hat{n}}
=0$ one finds the experimental value of the effective q.e. for the squeezed state at 568 nm. The value yields $\eta_{E,\rm SV}^{\rm exp}=0.17\pm 0.03$, and it was further used for the conversion of photocounts to photon numbers. Effective q.e. for the coherent state at 532 nm was found from $\eta_{E,\rm SV}^{\rm exp}$, assuming the dependence of q.e. on the wavelength, provided by the manufacturer, and yielding $\eta_{E,\rm coh}^{\rm exp}=0.20\pm 0.03$. 
%
The theoretical fits of 
the data to the model yield $N_{\rm max}=3$, $P=0.280\pm 0.005$, and $\eta=0.163\pm 0.007$ for the 
coherent state and $N_{\rm max}=3$, $P=0.30\pm 0.02$, and $\eta=0.145\pm 0.009$ for the SV state.  These 
parameters yield $\eta_{E,\rm coh}^{\rm fit}=0.21\pm 0.01$ and $\eta_{E,\rm SV}^{\rm fit}=0.19\pm 0.01$, which are in good agreement with the experimental values of $\eta_{E}$ used for the calculation of photon numbers. The results clearly 
demonstrate that NRF gradually decreases with the increasing $\braket{\hat{n}}$ for both 
studied states, which is caused by the saturation of the MPPC. At the same time, the NRF for 
the SV state lies below the one for the coherent state - a signature of squeezing. The 
crosstalk elevates the NRF at low $\braket{\hat{n}}$ to values above unity 
as predicted by the theory. Surprisingly, the found value of $N_{\rm max}$ appears to be much 
smaller than the one expected from the MPPC with 400 pixels. We have 
found that the MPPC modules do not allow reliable discrimination of photon numbers beyond
$\braket{\hat{n}}=5$ due to the limitations of the built-in electronic circuitry, which was inaccessible for any adjustment.
This might also be the cause of the deviation of the measured NRF from the theoretical prediction 
for $\braket{\hat{n}}>5$. Note that reliable photon 
number resolution up to 20 photons was demonstrated with similar detectors 
using customized electronics \cite{Eisenberg}.

\begin{figure}[htb]
\centerline{\includegraphics[width=7.5 cm]{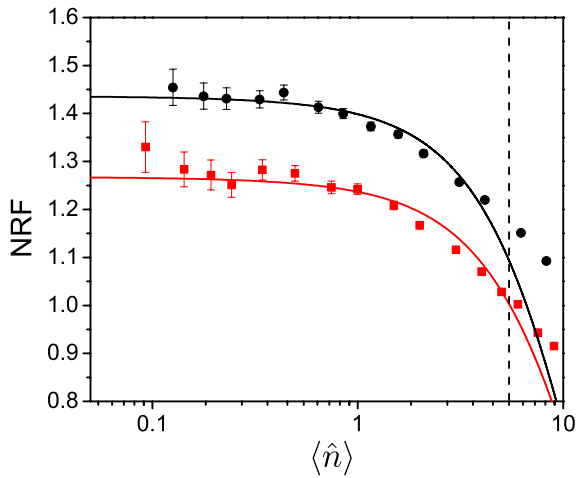}}
\caption{\small (Color online) Dependence of the NRF on the mean number of photons measured 
for the coherent state (black circles) and the SV state (red squares). The theoretical fits 
for the NRF are done for the range of average photon numbers up to 5 photons (vertical dashed 
line), which marks the range of reliable photon-number resolution. The solid lines are fits to 
the theoretical model with $R^2=0.9999$ and $R^2=0.9994$ for the coherent and SV states, respectively.}
\end{figure}

In conclusion, we have performed an experiment in which the NRFs of coherent and SV states 
have been measured with commercially available MPPC modules at relatively high photon numbers 
(up to 5). The main conclusions we can make are: (1) the 
crosstalk in the MPPC leads to an increase in the NRF for both coherent and SV states, (2) 
saturation of the MPPC leads to the decrease of NRF with increasing photon numbers. The 
experimental data agrees with the theoretical model which takes into account the saturation 
and the crosstalk although the fits deviate at higher mean photon numbers, where the photon 
number resolution of the detectors is limited. These results extend the use of MPPCs for the 
characterization of quantum light in a significantly broader range than that of conventional 
SPADs.

The work was supported by the A-STAR Investigatorship grant.

\pagebreak

\section*{Informational Fourth Page}

\end{document}